\documentclass[aps,pr,
showpacs,superscriptaddress,groupedaddress,nofootinbib]{revtex4}  

\usepackage{longtable}
\usepackage{graphicx}  
\usepackage{ulem}   
\usepackage{comment} 
\usepackage{datetime}
\usepackage{dcolumn}

\usepackage{amsxtra}
\usepackage{euscript} 
\usepackage{bm}        
\usepackage{tabularx}
\usepackage{float}
\restylefloat{table}

\usepackage[cmtip,arrow]{xy}
\usepackage{pb-diagram,pb-xy}

\usepackage{amsmath}
\usepackage{amssymb}
\usepackage{amsthm}
\usepackage[pdftex]{color}
\usepackage[pdftex,colorlinks,citecolor=blue,linkcolor=blue,urlcolor=blue]{hyperref} 
\usepackage{graphicx}
\usepackage{dcolumn} 
\usepackage{bm} 

\usepackage{longtable}

\usepackage{ulem}   
\usepackage{comment} 
\usepackage{datetime}

\usepackage{tabularx}
\usepackage{float}
\restylefloat{table}

\newcommand{\beq}{\begin{equation}}
\newcommand{\eeq}{\end{equation}}
\newcommand{\bea}{\begin{eqnarray}}
\newcommand{\eea}{\end{eqnarray}}
\newcommand{\barr}{\begin{array}}
\newcommand{\earr}{\end{array}}

\long\def\begincomment#1\endcomment{}

\newtheorem{theorem}{Theorem}

\usepackage{pgf,tikz}
\usetikzlibrary{arrows}

\usepackage{bm}
\usepackage{bbold}

\usepackage{mathtools}

\DeclarePairedDelimiterX\braket[2]{\langle}{\rangle}{#1 \delimsize\vert #2}

\begin{document}

\title{Low temperature dynamics for confined $p=2$ soft spin in the quenched regime
}

\author{Vincent Lahoche} \email{vincent.lahoche@cea.fr}
\affiliation{Université Paris Saclay, CEA List, Gif-sur-Yvette, F-91191, France}

\author{Dine Ousmane Samary}
\email{dine.ousmanesamary@cipma.uac.bj}
\affiliation{Université Paris Saclay, CEA List, Gif-sur-Yvette, F-91191, France}

\affiliation{International Chair in Mathematical Physics and Applications (ICMPA-UNESCO Chair), University of Abomey-Calavi,
072B.P.50, Cotonou, Republic of Benin}

\date{\today}

\begin{abstract}
This paper aims to address the low-temperature dynamics issue for the $p=2$ spin dynamics with confining potential, focusing especially on quartic and sextic cases. The dynamics are described by a Langevin equation for a real vector $q_i$ of size $N$, where disorder is materialized by a Wigner matrix and we especially investigate the self-consistent evolution equation for effective potential arising from self-averaging of the square length $a(t)\equiv \sum_i q_i^2(t)/N$ for large $N$. We first focus on the static case, assuming the system reached some equilibrium point, and we then investigate the way the system reaches this point dynamically. This allows identifying a critical temperature, above which the relaxation toward equilibrium follows an exponential law but below which it has infinite time life and corresponds to a power law decay.
\end{abstract}


\maketitle


\section{Introduction}

Glassy systems are usually characterized by their static properties, as replica symmetry breaking is the most famous example. Alternatively, they can be characterized by their dynamical aspects, and never reach equilibrium for experimental time scales below the ‘‘glass'' transition temperature. As the transition point is reached, relaxation time increases and the decay toward equilibrium becomes slower than exponential law \cite{Dominicsbook}.  The soft $p$-spin model is a popular mathematical model that allows describing such a glassy system \cite{Leticia1}-\cite{Rokni} see also \cite{Mezard1,Caiazzo} and references therein. It describes the dynamics of $N$ random variables $q_i\in \mathbb{R}$ through a Langevin-like equation where disorder is materialized by a rank $p$ random real and symmetric tensor $J_{i_1i_2\cdots i_p}$:
\begin{equation}
\frac{d q_i}{dt}=- \frac{\partial}{\partial q_i} V_J[q(t)]-\ell(t) q_i(t) +\eta_i(t)\,,\label{langevin1}
\end{equation}
 where the interaction potential $V_J[q(t)]$ in the Langevin equation involves the coupling tensor $J_{i_1i_2\cdots i_p}$ and is given by:
\begin{equation}
V_J[q]:=\frac{1}{p}\sum_{i_1,\cdots,i_p} J_{i_1 i_2\cdots i_p}\, q_{i_1}\cdots q_{i_p}\,,
\end{equation}
$\eta_i(t)$ is a Gaussian random field with Dirac delta correlations:
\begin{equation}
\langle \eta_i(t) \eta_j(t^\prime) \rangle = 2T \delta_{ij} \delta(t-t^\prime)\,,\label{distributioneta}
\end{equation}
and the function $\ell(t)$ avoids large values configurations for $q_i$'s. The parameter $T$ involved in the definition (\ref{distributioneta}) identifies physically as the temperature regarding the equilibrium states. In this paper, we focus on the case $p=2$, where the disorder is represented by a Gaussian random matrix with variance $\sigma^2$. For the spherical model, $\sum_{i=1}^N q_i^2=N$, and $\ell(t)$ is a Lagrange multiplier. Alternatively, $\ell(t)$ can be a $O(N)$ invariant polynomial function: $\ell(t):=2\sigma+\sum_n h_n a^n(t)$ with $a(t):=\sum_i \frac{q_i^2(t)}{N}$. The function $\ell(t)$ derives from a \textit{potential} $\mathcal{V}(a)$ as
\begin{equation}
\ell-2\sigma=:\frac{\partial \mathcal{V}}{\partial a}\,,
\end{equation}
in which the extra-term $2\sigma$ is chosen for convenience, see the section \eqref{sec2} for more detail and so that the right-hand side of the Langevin equation \eqref{langevin1} looks as a gradient flow \cite{Kristima,Altieri}.
\medskip

In the large $N$ limit, the ‘‘hard sphere" spherical $p=2$ spin dynamics has been investigated analytically twenty-five years ago \cite{Leticia1}, exploiting Wigner semi-circle law for the eigenvalue distribution of the disorder $J_{ij}$. As a result, even though the $p=2$ spherical spin glass looks like a ferromagnetic in disguise \cite{Dominicsbook,Kristima} rather than a true spin glass regarding its statics properties\footnote{In particular, no replica symmetry breaking occurs.}, its dynamics is however non-trivial. Indeed below the critical temperature $T_c$, the system never reaches equilibrium with exponential decay except for very special initial ‘‘staggered'' configurations for $q_i(t=0)$'s and ergodicity is weakly broken. As for the static limit, this behavior is reminiscent of the domain coarsening for a ferromagnet in the low-temperature phase \cite{Dominicsbook,Kristina1}, where equilibrium fails as the size of the domains with positive and negative magnetization grows in time. . 
\medskip

 This paper is a companion of the reference \cite{Vincent}, aiming to address specifically the issue to solve asymptotic dynamics from a low-temperature expansion in the large $N$ regime, in the case where the sharp spherical constraint is replaced by a smooth potential with some stable minima. We suggest a general formalism to investigate the asymptotic equation satisfied by $a(t)$ and $\dot{a}(t)$ arising because of the self-averaging of $a(t)$. This equation turns to be a closed equation for $a(t)$ assuming to remains close to some equilibrium point of the potential for late time, and can be easily solved by Laplace transform accordingly to the method considered in \cite{Bray,Emmot} for phase ordering kinetics. The solution exhibits an explicit expression for the critical temperature toward a weak ergodicity breaking phase, valid around all equilibrium points of the potential. Analysing the way the system reaches an equilibrium point, we show that the hypothesis depends strongly on the degeneracy of the vacuum. 
\medskip

The reference \cite{Vincent}, and the present paper are both in the continuation of \cite{Lahochepspin} and in the continuity of our series of paper \cite{Lahochesignal0,Lahochesignal1,Lahochesignal2,Lahochesignal3,LahochesignaREV}. Indeed in these works, we investigated the detection signal issue for nearly continuous empirical spectra from the point of view of an effective field theory approaches, and the model considered in this paper and \cite{Vincent} can be viewed as an out-of-equilibrium version of this approaches until now focused on equilibrium theory that corresponds indeed to equilibrium states of the model considered in this work. We consider indeed such an out-of-equilibrium model for signal detection in a work to appear \cite{prepa}, and it should be noticed that the field theory underlying this analysis is also close to the spin glass models considered in \cite{Sompolinsky1,Sompolinsky1}. 
\medskip

Besides the bibliographic line of the authors, this work finds a place in the general context of the dynamics of glassy systems. It has been considered in \cite{Kristima} for the quartic version, and for instance in \cite{Guionnet,Sompolinsky1,Sompolinsky2,Kurchan}. For long-time physics, this point of view is expected equivalent to the p-spin dynamics with sharp spherical or hard spins constraints \cite{Dominicsbook,Leticia1,Kristima,Kristina1,Nishimori,Cugliandolo1,Cugliandolo2,Cugliandolo3}, provided that the minimum of the potential is non-degenerate, that we consider in our work. Furthermore, as pointed out in \cite{Annibale},  
 taking into account fluctuations of extensive quantities requires one to be careful in sharp spherical models. Finally, the underlying physics for materials for $p$ spin models and especially aging effects are discussed in \cite{vincent2,herisson,Cugliandolo4}. Moreover, because of its links with ferromagnet and coarsening phenomenon \cite{Dominicsbook}, phenomenology is also close to phase ordering kinetic literature \cite{Bray,Emmot,Livi}.
\medskip

\noindent
\textbf{Outline:} The paper is organized as follows. In section \ref{sec2} we consider the late-time closed equations arising in the quenched regime from the assumption that the system stays around some equilibrium point of the potential, and we consider the sextic case as an illustration. In section \ref{conv} we investigate the way the system converges toward the equilibrium point. In particular, we show that it depends on the degeneracy of the vacua. In particular, for a doubly degenerate vacuum, we show that the system indeed relaxes exponentially toward the zero vacuum for $T$ small enough. Concluding remarks are found in section \ref{sec6}.


\section{Late time closed equation in the quenched regime}\label{sec2}

\paragraph{The closed equations.} The disorder matrix $J_{ij}$ is a real symmetric matrix and can be diagonalized with eigenvalues $\lambda \in \mathbb{R}$.
Into the eigenspace, the Langevin equation (\ref{langevin1}) for $p=2$ reads:
\begin{equation}
\frac{d q_\lambda}{dt}=-[\lambda+\ell(t)]q_\lambda(t)+\eta_\lambda(t)\,,\label{langevin2}
\end{equation}
where $q_\lambda:=\sum_i q_i u_i^{(\lambda)}$ is the projection along the eigenvector $u_i^{(\lambda)}$. For large $N$, eigenvalues $\lambda$ are assumed to display accordingly with the Wigner semi-circle law \cite{Potters} with variance $\sigma^2$:
\begin{equation}
\frac{1}{N}\sum_\lambda f(\lambda) \to \int_{-2\sigma}^{2\sigma} \,\mu(\lambda) f(\lambda) d\lambda\,,
\end{equation}
where $\mu(\lambda)$ is the standard Wigner semicircle distribution:
\begin{equation}
\mu(\lambda):=\frac{\sqrt{4\sigma^2-\lambda^2}}{2\pi \sigma^2}\,.
\end{equation}
Equation (\ref{langevin2}) can be solved formally taking $t=0$ as the initial condition:
\bea
q_\lambda(t)=q_\lambda(0)\,e^{-(2\sigma + \lambda)t}\rho(t)+\int_0^t dt^\prime\, e^{-(2\sigma+\lambda)(t-t^\prime)}\,\eta_\lambda(t^\prime)\frac{\rho(t)}{\rho(t^\prime)}\,,\label{evolution}
\eea
with:
\begin{equation}
\rho(t):=e^{2\sigma t - \int_0^t dt^\prime \ell(t^\prime)}\,.
\end{equation}
We are aiming to investigate the large-time behavior of the Langevin equation (\ref{langevin2}), focusing on the function $g(t)$ defined as:
\begin{equation}
g(t)=\int_0^t\, dt' \ell(t')\,.
\end{equation}
Assuming uniform initial condition for $q_\lambda$, namely\footnote{More precisely, we can show uniform configuration $q_\lambda(0)=c$ not too far from one of the local minima of the potential.} $q_\lambda(0)=1\,\forall \lambda$\footnote{Physically, this condition is equivalent to assume that variables $q_i$ are randomly distributed for $t=0$.}, we get after the quench for the expectation value of $q_\lambda^2(t)$:
\bea\label{xxmodel}
\langle q_\lambda^2\rangle=e^{-2g(t)}\left[e^{-2\lambda t}+2T\int_0^t dt'\, e^{-2\lambda(t-t')+2g(t')}\right]\,,
\eea
where $a(t)$ involved in this equation is assumed to be self-averaged. Because:
\begin{equation}
a(t)=\int_{-2\sigma}^{2\sigma} d\lambda\,\mu(\lambda) \langle q_\lambda^2\rangle\,,\label{a(t)}
\end{equation}
the previous equation leads to a formally closed equation for $a(t)$:
\begin{equation}
\boxed{a(t)=G^{-1}(t) \left[ H(t)+2T F(t) \right] }
\end{equation}
where:
\bea\label{momi}
G(t):=\exp \Big(2g(t)-4\sigma t\Big),\quad
H(t):=\int_{-2\sigma}^{2\sigma}\mu(\lambda) e^{-2\lambda t-4\sigma t}d\lambda.\label{defGH}
\eea
and where $F(t)$ is the convolution of the above function $G(t)$ and $H(t)$ i.e.:
\begin{equation}
F(t)=\int_0^t dt'\, H(t-t')G(t')\,.\label{defF}
\end{equation}
Note that the function $H(t)$ can be computed exactly, and we get:
\bea
H(t)=\frac{e^{-4\sigma t}\mathcal{I}_1(4\sigma t)}{2\sigma t}\,,
\eea
where $\mathcal{I}_n(x)$ is the standard modified first kind of Bessel function. Explicitly, for $\sigma=1$ and $t$ large enough:
\bea
H(t){\approx}\frac{1}{4\sqrt{2\pi}}\frac{1}{t^{\frac{3}{2}}}\,.
\eea
One can obtain a second and more tractable closed equation for $G(t)$ from the observation that thermal fluctuations must have an effect to precipitate the system on the equilibrium point of the potential, namely the points where
\begin{equation}
\ell(t)-2\sigma:= h_0+\sum_{n=1}^K h_n a^n(t)\,,
\end{equation}
vanishes. The consistency of this assumption will be investigated further in this paper, but for this section, we assume that for time large enough, $\ell(t) \to 2\sigma$. To be valid for all times, the equation $\ell(t) -2\sigma=0$ requires:
\begin{equation}
a(t)=\gamma\,,
\end{equation}
where $\gamma$ is a real and positive number such that:
\begin{equation}
h_0+\sum_{n=1}^K h_n \gamma^n=0\,.
\end{equation}
Obviously, there is not a single solution in general, but with the deep of the wells being large in the large $N$ limit, this conclusion agrees with the intuition that the system around one of the minimum of the potential is blinded from the other minima. From the explicit expression for $a(t)$ given by equation \eqref{a(t)} leads to:
\begin{equation}
\boxed{G(t)=\frac{1}{\gamma}H(t)+\frac{2T}{\gamma} \,F(t)\,.} \label{closedTWO}
\end{equation}

\paragraph{Formal solution.} 

 Equation of the form \eqref{closedTWO} is similar to the closed equation arising for the spherical case, which has been mainly addressed in the literature -- see for instance \cite{Dominicsbook} and reference therein for a detailed treatment of the spherical model. The main difference for the confining potential case is that the closed equation is only an asymptotic relation, valid for late times, whereas it holds for all time in spherical dynamics. However, for $t$ large enough, one can expect that $H(t-t^\prime)$, which behaves as $1/t^{3/2}$, suppresses low-time contributions for $G(t)$, provided it has a finite limit for short times and that it decreases slowly enough. Furthermore, one may expect fluctuations of $\ell(t)$ to have a small standard deviation around the large time average in the large $N$ limit. In that way, it is reasonable to assume that the solution of the closed equation \eqref{closedTWO} (for \textit{all times}) provides us with the true asymptotic behavior for $G(t)$ in the late time limit. This assumption has been done in Appendix B of \cite{Vincent} and we shortly review the method in this section. \\

\begin{figure}
\begin{center}
\includegraphics[scale=0.4]{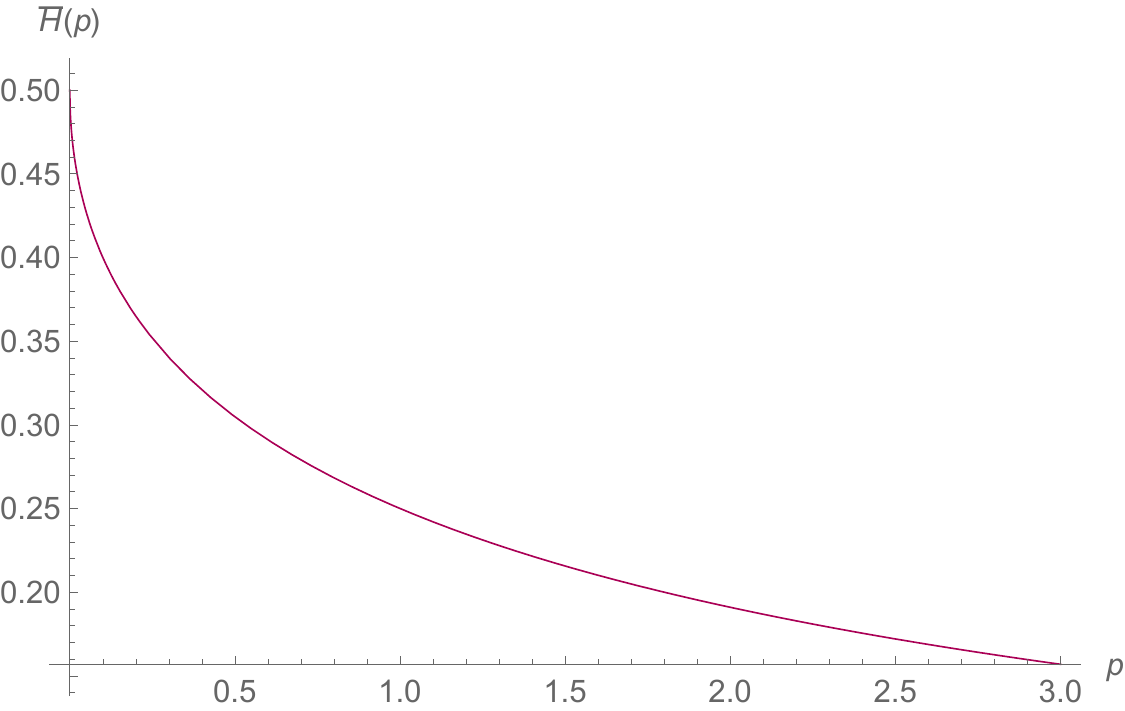}
\end{center}
\caption{Typical shape of the function $\bar{H}(p)$, for $\sigma=1$.}\label{figH}
\end{figure}

The closed equation \eqref{closedTWO} can be easily solved using standard Laplace transform methods. For some function $f(t)$, the Laplace transform (if it exists), is defined as:
\begin{equation}
\bar f(p):=\int_0^\infty\, dt\, e^{-pt}f(t)\,.
\end{equation}
Hence, the solution of the closed equation \eqref{closedTWO} reads formally as:
\begin{equation}
\bar G(p)=\frac{1/2}{\frac{1}{2}\gamma\bar H^{-1}(p)-T}\,,
\end{equation}
where the Laplace transform of $H(t)$ reads explicitly:
\bea
\bar H(p)=\frac{4\sigma+p-\sqrt{(4\sigma+p)^2-16\sigma^2}}{8\sigma^2}\,,\label{H(p)}
\eea
and Figure \ref{figH} shows the typical shape of $\bar{H}(p)$ for $\sigma=1$. Because the function $\bar G(p)$ has to be positive, we must have $\gamma >0$. Furthermore, the temperature $T$ has to be smaller than the critical temperature $T_c$ defined as:
\begin{equation}
\boxed{T<T_c:=\frac{1}{2}\gamma\bar H^{-1}(0)\equiv\gamma \sigma\,.}
\end{equation}
At the critical temperature, the expression for $G(p)$ becomes singular, and the critical temperature is nothing but the radius of convergence of the power series expansion in $T$ for ${G}(t)$. To understand the large-time behavior of $G(t)$, we expand $H(p)$ around $p=0$. From the explicit expression \eqref{H(p)}, we get:
\bea
\bar H(p)\approx \bar{H}(0)-\frac{\sqrt{\sigma}}{4\sqrt{2}}\,p^{1/2}\,,
\eea
and:
\bea
\bar G(p)\approx \frac{1}{2(T_c-T)}\Big[1-\frac{\tilde{A}(\sigma)T_c^2 p^{1/2}}{T_c-T}+\mathcal{O}(p)\Big]\,,\label{staticG}
\eea
where $\tilde A(\sigma):=\frac{\sqrt{\sigma}}{2\gamma\sqrt{2}}$.
The asymptotic expression for $G(t)$ can be obtained, for small $p$, from standard results about the asymptotic expression of inverse Laplace transforms near the origin. In \cite{Richard} for instance, one can find the following statement:
\begin{theorem}\label{theorem1}
Let $f(t)$ be a locally integrable function on $[0,\infty)$ such that $f(t)\approx \sum_{m=0}^\infty c_m t^{r_m}$ as $t\rightarrow\infty$ where $r_m<0$. If the Mellin transformation of this function is defined and if no $r_m=-1,-2,\cdots$ then the Laplace transformation of $f(t)$ is
\beq
\bar f(p)=\sum_{m=0}^\infty c_m \Gamma(r_m+1)p^{-r_m-1}+
\sum_{n=0}^\infty Mf(n+1)\frac{(-p)^n}{n!}
\eeq
where
$Mf(z)=\int_0^\infty\, t^{z-1}f(t)dt$ is the Mellin transform of the function $f(t)$.
\end{theorem}
Hence, for $t$ large enough we expect that $G(t)\sim t^{-3/2}$, implying: $\ell(t) \sim 2\sigma+ \mathcal{O}(t^{-1})$: the relaxation of the system below the critical temperature behaves as a power rather than an exponential i.e. has infinite relaxation time. Furthermore, the late time $2$-points correlation $C(t)$ defined as:
\begin{equation}
C(t):=\int_{-2\sigma}^{2\sigma} d\lambda\, \mu(\lambda) \langle q_\lambda(t) q_\lambda(0) \rangle\,,
\end{equation}
behaves as:
\begin{equation}
C(t)\sim \frac{1}{t^{3/4}}\,,
\end{equation}
and the memory of the initial condition is long i.e. it has infinite exponential time life. The previous method breaks down for high temperatures, and we expect this is a consequence of the fact that $G(t)$ diverges faster than any power law above $T_c$, such that Laplace transform for arbitrary small $p$ does not exist. Hence, above $T_c$ the system is expected to relax toward equilibrium accordingly with an exponential law, and the memory of the initial condition has a finite time life.

\begin{figure}
\begin{center}
\includegraphics[scale=0.3]{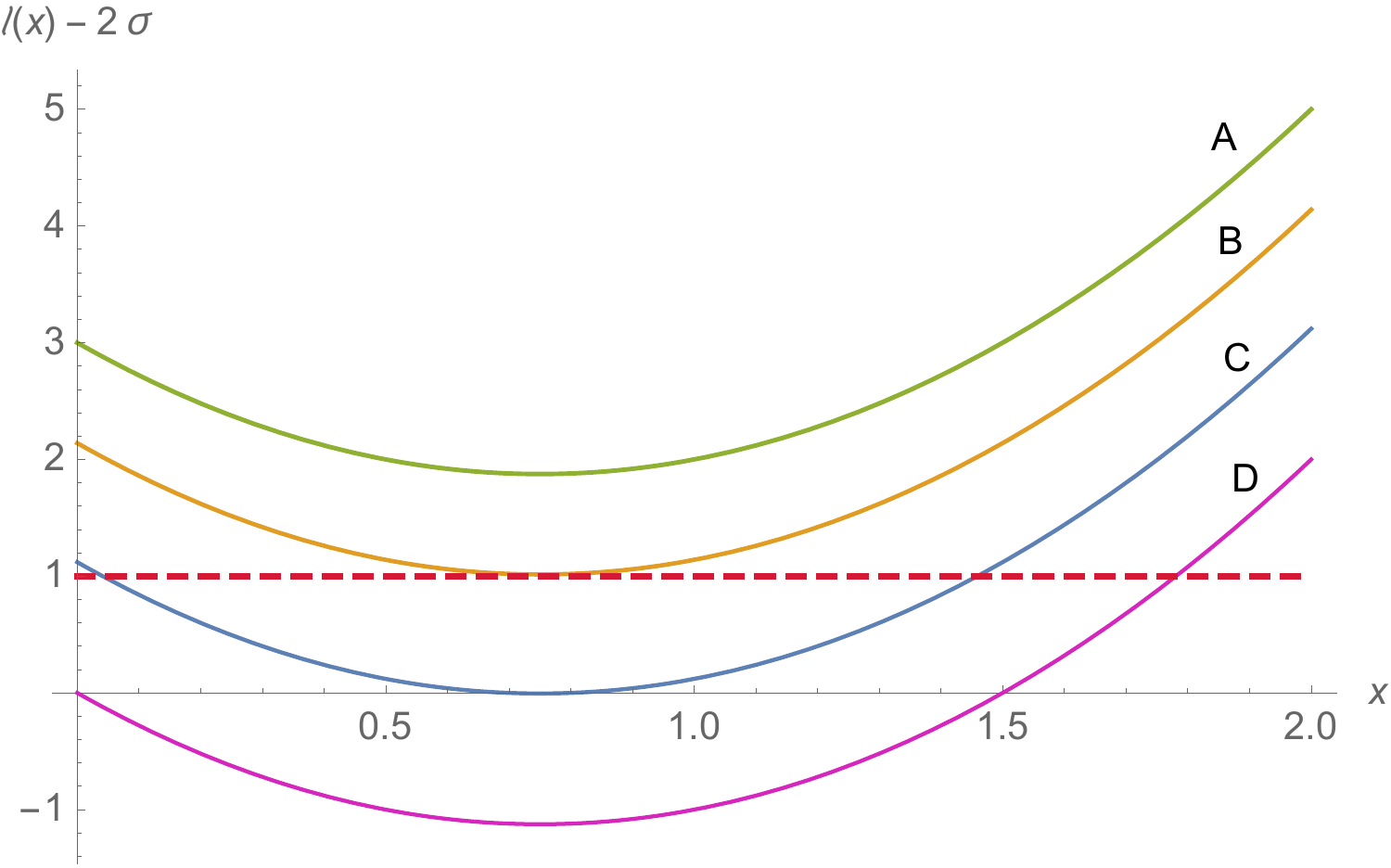}\quad\includegraphics[scale=0.3]{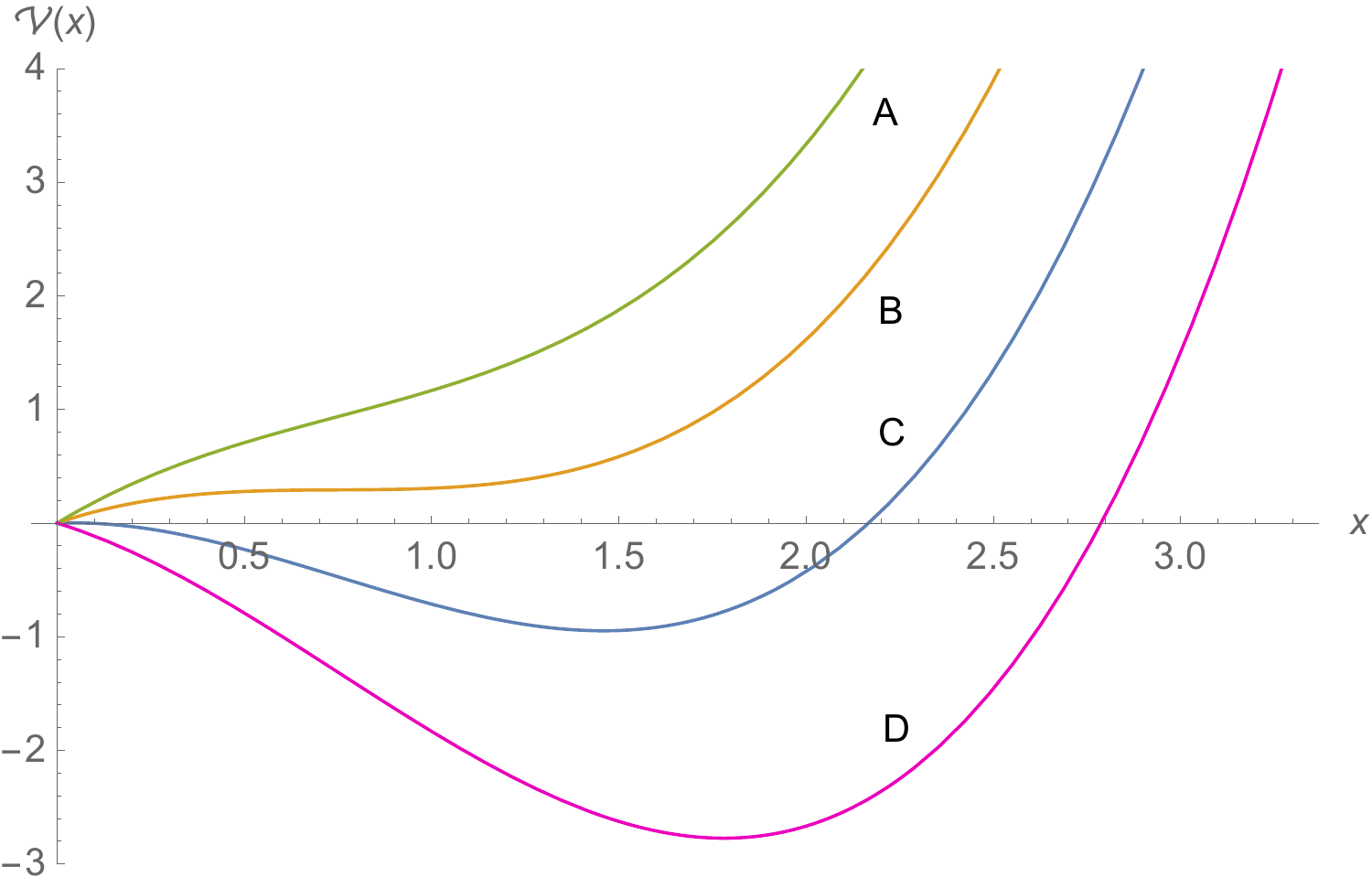}
\end{center}
\caption{Solutions for $\gamma_{\pm}$ regarding the shape of the potential $\mathcal{V}(x)$ ($\sigma=1/2$).}\label{figfinal}
\end{figure}

\paragraph{Example with a sextic potential.} As an illustration, let us investigate the case of a sextic potential (regarding the power of the variables $q_i$):
\begin{equation}
\mathcal{V}(x):=\frac{1}{3} h_2 x^3+\frac{1}{2}h_1x^2+h_0 x\,,
\end{equation}
Furthermore, the equation for $\gamma$ is nothing but:
\begin{equation}
\ell(\gamma_\pm)-2\sigma:=\frac{\partial \mathcal{V}}{\partial x}(\gamma_\pm)\equiv 0\,,
\end{equation}
and the two nonzero solutions are explicit:
\begin{equation}
\gamma_\pm=-\frac{h_1}{2h_2}\pm \frac{\sqrt{h_1^2-4 h_2 h_0}}{2 h_2}\,.
\end{equation}
The coupling $h_2$ must be positive because of the stability requirement. Hence, if $h_1>0$, we have two configurations:
\begin{itemize}
\item For $h_0<0$, there is a single positive solution, $\gamma_+$.
\item For $h_0>0$, the two solutions are negative or imaginary, and the low $T$ expansion does not exist.
\end{itemize}
For $h_1<0$ on the other hands,
\begin{itemize}
\item For $h_0>0$, as soon as $h_1^2>4h_2h_0$, there are two solutions for $\gamma_\pm$. There are no solutions for $h_1^2<4h_2h_0$.
\item For $h_0<0$ finally, there are only one solution again, namely $\gamma_+$.
\end{itemize}
One can found on Figure \ref{figfinal} and illustration of this behavior for $h_1<0\,(=-3)$ and $h_2=1$, setting $\sigma=0.5$. For curves labeled from $(A)$ to $(D)$, the value of the quadratic term $h_0$ decreases. Especially for the curve labeled by $(A)$, $h_0>0$ and the discriminant is negative, and as a result, the solution set is empty ( the potential has a single minimum located for $x=0$). As $h_0$ decreases in $(B)$ to $(C)$, the discriminant becomes positive and two solutions appear. But as $h_0<0$, the first solution ultimately disappears, which leads us to a single solution, as the transition regime with phase coexistence is over.

\section{Convergence toward equilibrium points}\label{conv}

In this section, we investigate the validity of the previous assumptions, regarding especially the convergence toward the equilibrium point for large $t$, that in particular will confirm the validity of the expected late-time behavior for $G(t)$. Assuming the validity of the quenched limit, and because of the definition of $G(t)$, we have $\dot{G}(t)=2G(t) (\ell(t)-2\sigma)$, or explicitly:
\begin{equation}
\dot{G}(t)=2h_KG(t)\prod_{\mu=1}^R \left(\frac{H(t)+2T F(t)}{G(t)}-\gamma_\mu \right)^{r_\mu}\,,\label{equationdymgen}
\end{equation}
where we set $h_K=1/2$ for simplicity, $R$ designates the number of zeros and $r_\mu$ their multiplicity. These are the equations that we will consider in this section. \\

\begin{figure}
\begin{center}
\includegraphics[scale=0.65]{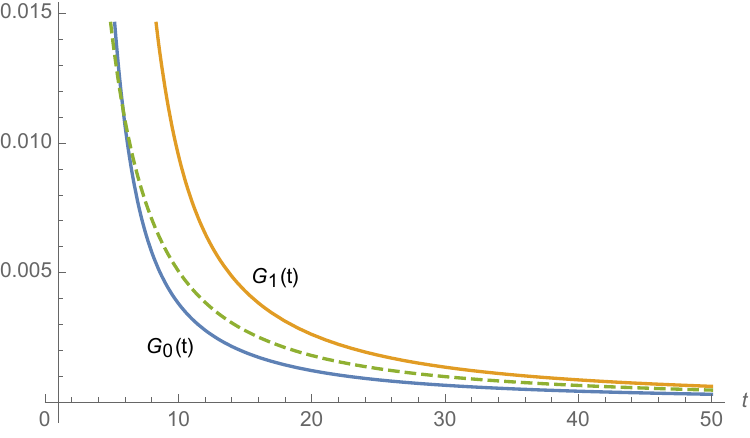}\,\,\,\,
\includegraphics[scale=0.65]{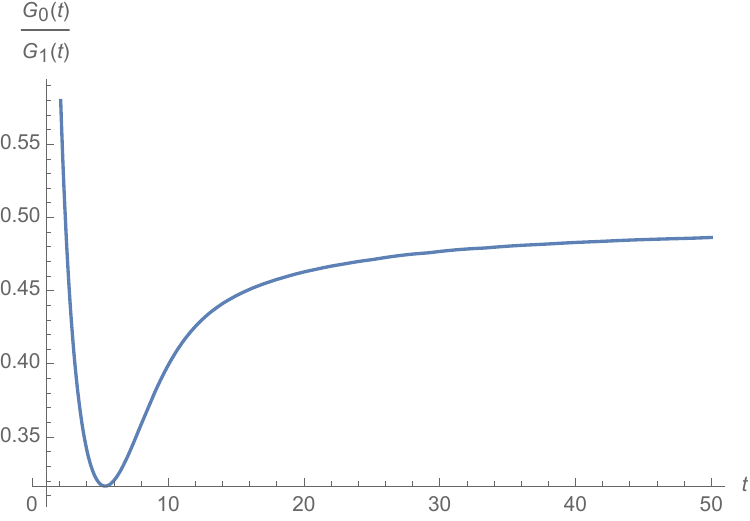}
\end{center}
\caption{On the top: 
 Typical behavior of $G_0(t)$ (blue curve) and $G_1(t)$ (yellow curve) for large time, the green dotted curve is $f(t):=0.16/t^{3/2}$. On the bottom: the behavior of $G_0(t)/G_1(t)$, which goes toward a constant for a large time.}\label{figplot1}
\end{figure}

\paragraph{Solution for quartic potential.} For a quartic potential, assuming $\gamma >0$ and setting $h_1=1/2$, the effective equation of motion for $G(t)$ reads:
\begin{equation}
\dot{G}(t)=-\gamma G(t) + H(t)+2T F(t)\,.
\end{equation}
Once again, this equation can be solved using elementary Laplace transform techniques, and because of the initial condition $G(0)=1$, we get:
\begin{equation}
\bar{G}(p)=\frac{1+\bar{H}(p)}{p+\gamma-2T\bar{H}(p)}\,.
\end{equation}
For small values of $p$, we find that the large-time dynamics is still dominated by the behavior of $\bar{H}(p)$, and the leading order contribution for $\bar{G}(p)$ is a slightly modified version of \eqref{staticG}:
\begin{equation}
\bar{G}(p)=\frac{1/2}{T_c-T} \left[1+2\sigma-\frac{A(\sigma)T_c^2}{T_c-T}\left(1+\frac{2T}{\gamma}\right)\sqrt{p}\right]\,,
\end{equation}
and again $G(t)\sim t^{-3/2}$ for late times (Theorem \ref{theorem1}). Note that the critical temperature $T_c\equiv\gamma\sigma$ has the same value as in the static limit. An understanding of global dynamics should require numerical methods. Focusing on the small temperature regime, one can make an expansion of the form:
\begin{equation}
G(t)=\sum_{n=0}^\infty\, (2T)^n G_n(t)\,,
\end{equation}
and explicitly:
\begin{equation}
G_0(t)=e^{-\gamma t}+F_0(t),\quad 
G_1(t)=\int_0^t G_0(t-t^\prime) F_0(t^\prime)\,,
\end{equation}
with:
\begin{equation}
F_0(t):=\int_0^t H(t-t^\prime) e^{-\gamma t^\prime}\,,
\end{equation}
and for $n>0$ we have the general recursive relation:
\begin{equation}
G_n(t)=\int_0^t G_{n-1}(t-t^\prime) F_0(t^\prime)\,.
\end{equation}
Figure \ref{figplot1} shows the behavior of $G_0(t)$ and $G_1(t)$ for $\sigma=\gamma=1$. Both have a maximum after which they decrease to $0$ as $t^{-3/2}$. \\

\paragraph{Higher potentials and degeneracy.} The case of a general potential having many minima could be difficult to investigate with elementary methods, except if we assume that the system is initially close to one of the isolated minimums, say $\nu$. In that case, one expects, because of the large depth of the well that the system remains close to this minimum during its dynamics. This can be achieved for instance by replacing our initial condition $q_\lambda(0)=1$ by $q_\lambda(0)=\sqrt{\Delta}$, such that $a(0)=\Delta$ and $\vert \Delta- \gamma_\nu \vert \ll 1$. Formally, assuming we remain close to the minimum $\gamma_\nu$ (that has to be checked to be self-consistent) i.e. $\vert a(t)-\gamma_\mu \vert \equiv \epsilon(t) \ll 1$, and that spacing between minima is large enough, equation \eqref{equationdymgen} rewrites as:
\begin{equation}
\frac{\dot{G}}{G}= 2h_K R(\gamma_\nu) \left( \frac{\Delta H(t)+2T F(t)}{G(t)}-\gamma_\nu\right)^{r_\nu}+\mathcal{O}(\epsilon^{r_\nu+1})\,,\end{equation}
where $R(\gamma_\nu)$ denotes the remaining of the euclidean division of $\ell(t)-2\sigma$ by $(a(t)-\gamma_\nu)^{r_\nu}$, evaluated at the point $a(t)=\gamma_\nu$, and we can set $2h_K R(\gamma_\nu)=1$ such that, close to the zero $\gamma_\nu$:
\begin{equation}
\dot{G} \approx G(t) \left( \frac{\Delta H(t)+2T F(t)}{G(t)}-\gamma_\mu\right)^{r_\nu}\,.
\end{equation}
For $r_\nu=1$, except for the $\Delta$ in front of $H(t)$, we recover exactly what we obtained for the quartic case, and we investigate the case $r_\nu >1$ in the rest of this paragraph. \\

\begin{figure}
\begin{center}
\includegraphics[scale=0.65]{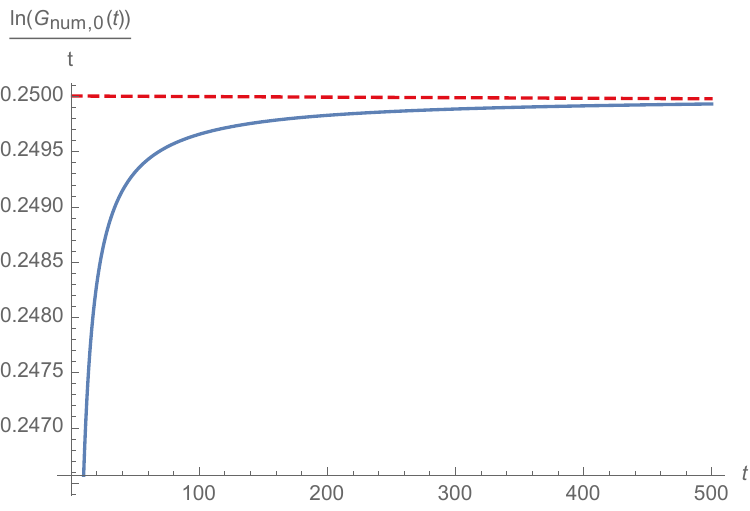}\quad
\includegraphics[scale=0.65]{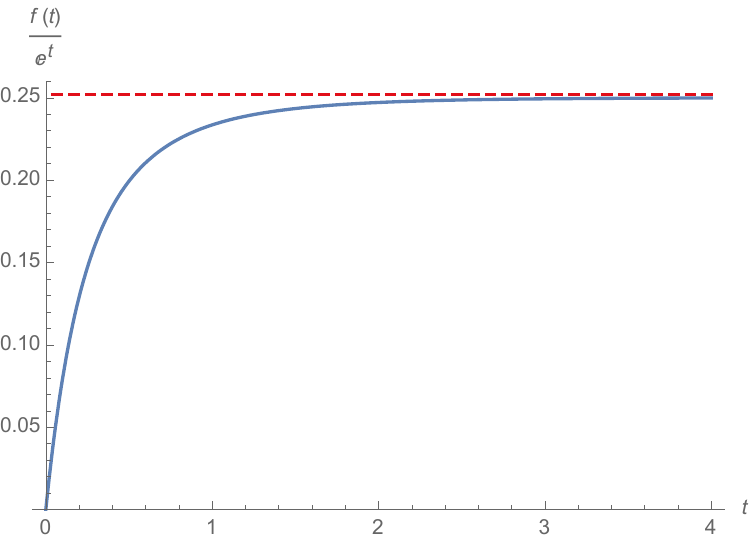}
\end{center}
\caption{On the top: Typical numerical behavior of $G(t)$ at $T=0$ for large $t$ with $\sigma=1$ and $\gamma=0.5$. On the bottom, behavior of $f(t)e^{-t}$ for $\sigma=1$ and $a^2=1$.}\label{figplot2}
\end{figure}

The behavior of the system will be different for odd or even values of $r_\mu$. Let us consider the case of an even value, and especially the case $r_\mu=2$:
\begin{equation}
\dot{G}=G(t) (a(t)-\gamma)^2\,.
\end{equation}
In that case, because $G(t)$ is a positive function, $\dot{G} > 0$ and the function $G(t)$ increases for all times. For $T=0$, it is easy to check that $G(t)=e^{\alpha t}$ is an asymptotic solution for the dynamics provided that $\alpha=\gamma^2$. This can be checked numerically, and the  typical late-time behavior of the system is pictured in Figure \ref{figplot2} ($G_{\text{num}}(t)$ denote the numerical solution of the equation) for some values of numerical parameters\footnote{Let us emphasize  that the figures illustrate the typical behavior of the system, and similar curves are obtained with different values of the parameters. }. If we assume the system converges to some finite value $a_\infty$, we must have $(a_\infty-\gamma)^2=\gamma^2$. The condition $a_\infty=0$ that solves this equation means the system goes toward the zero “vacuum” because of the disorder. \\
\medskip

\begin{figure}
\begin{center}
\includegraphics[scale=0.65]{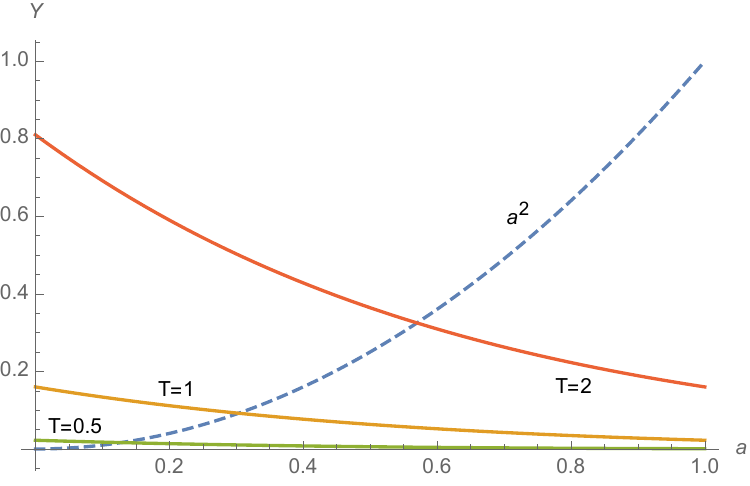}\,\,\,\,\includegraphics[scale=0.65]{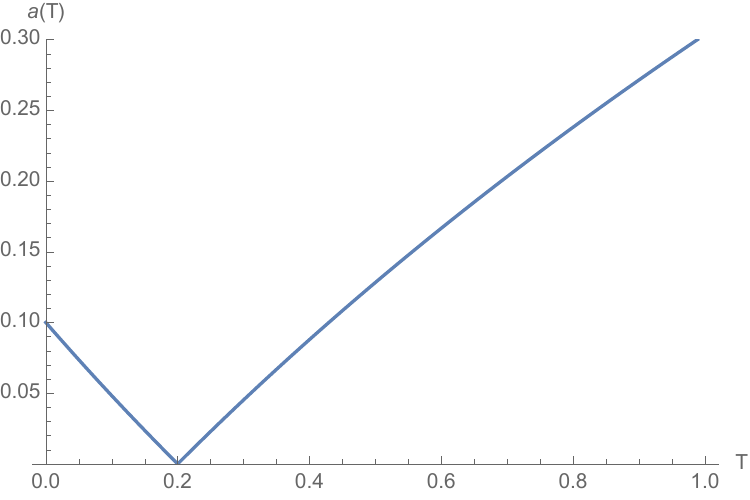}
\end{center}
\caption{On the top: Graphical solution of the equation for $\sigma=1$ and $\gamma=0.1$. On the bottom: explicit dependency of $a(T)$ on the temperature for $\gamma=0.1$.}\label{figplot3}
\end{figure}

For small $T$, one expects that the exponential behavior holds, and indeed:
\begin{equation}
f(t):= \int_0^t H(t-t^\prime) e^{a^2 t^\prime} \sim Q(a^2,\sigma) e^{a^2 t}\,,
\end{equation}
where the function $Q(a^2,\sigma)$ is explicitly:
\begin{equation}
Q(a^2,\sigma):=\frac{-a \sqrt{a^2+8 \sigma }+a^2+4 \sigma }{8 \sigma ^2}\,.
\end{equation}
Hence, asymptotically, the equilibrium value for $a^2$ is given by the solutions of the equation :
\begin{equation}
a^2=(2T Q(a^2,\sigma) - \gamma)^2=:Y(a,\sigma,\gamma,T)\,,
\end{equation}
 and the typical dependency on $T$ of the solutions is illustrated on Figure \ref{figplot3}, for fixed values of $\gamma$ and $\sigma$. The existence of a transition temperature that signals the end of the exponential regime can be proved by the following argument. Let $G(t)=\mathcal{A}(t) e^{a^2_* t}$ such that $\mathcal{A}(0)=1$ and $a^2_*$ a solution of equation $a^2=Y(a,\sigma,\gamma,T)$. Assuming $a(t)\ll 1$ accordingly with our zero temperature analysis, we have:
\begin{equation}
\dot{\mathcal{A}}\simeq -2\gamma \mathcal{A}(t) a(t) +\mathcal{A}(t) \delta \gamma^2\,,
\end{equation}
where $\delta \gamma^2:=\gamma^2-a^2_*$. Introducing $\mathcal{B}(t):=\mathcal{A}(t) e^{-\delta \gamma^2 t}$, the effective equation for $\mathcal{B}(t)$ can be  solved by Laplace transform; and we get:
\begin{equation}
\bar{\mathcal{B}}(p)\simeq\frac{1-2\gamma \Delta \bar{H}(p+\gamma^2)}{p+4\gamma T \bar{H}(p+\gamma^2)}\,.
\end{equation}
Hence, for large $t$, one expect the following behavior:
\begin{equation}
\mathcal{B}(t) \sim e^{-4\gamma T \bar{H}(\gamma^2)t}\,,
\end{equation}
and $G(t)$ increases exponentially provided that:
\begin{equation}
T< T_c\simeq \frac{\gamma}{4\bar{H}(\gamma^2)}\,.
\end{equation}

\begin{figure}
\begin{center}
\includegraphics[scale=0.65]{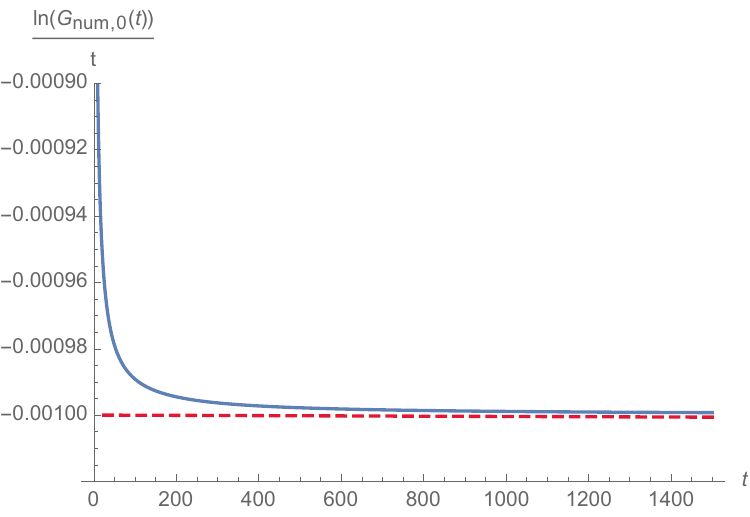}
\end{center}
\caption{{\color{blue}Typical} behavior of $G_{\text{num}}(t)$ for late time, with $\sigma=1$ and $\Delta=\gamma=0.1$}\label{figplot4}
\end{figure}

To conclude, let us investigate the case of an odd value for $r_\mu$, focusing on $r_\mu=3$ for numerical study. The typical behavior of $G_{\text{num}}(t)$ for $T=0$ is shown in Figure \ref{figplot4}, assuming $\Delta$ such that the initial condition is close enough to the degenerate vacuum. Numerically, one finds that the system behaves as $G_{\text{num}}(t) \sim e^{-\gamma^3 t}$ for late time. Once again, a transition temperature can be estimated by the following argument. The equilibrium equation $(a(t)-\gamma)^3=-\gamma^3$ admits in particular the unstable zero $a=0$. Hence, assuming $a(t) \sim 0$, the equation can be linearized, and one finds that the system goes toward the non-zero vacuum $a=\gamma$ as $1/t$, provided that $T< T_c = \gamma^3/2\bar{H}(0)$.

\section{Summary and outlooks}\label{sec6}

In this paper, we investigated the low-temperature behavior of the asymptotic quenched dynamics for a $p=2$ soft spin model with polynomial confining potential. We showed that a closed equation for $a(t)$ arises in the quenched limit, which is difficult to solve, and the first part of our analysis was devoted to a systematic analysis of the asymptotic closed equation for $G(t)$ arising from the assumption that the system relaxes toward one of the equilibrium points of the potential. The assumption that the solution of this equation provides the true asymptotic behavior for the function $G(t)$ is consistent only below a critical value $T<T_c$ for the temperature, and in this regime, the closed equation can be solved using Laplace transform methods. This leads to a power law $G(t)\sim t^{-3/2}$, and $\ell(t) \sim 2\sigma + \mathcal{O}(t^{-1})$. Besides the assumption that the system has small fluctuations around some zero of the function $\ell(t)-2\sigma$ and that $H(t-t^\prime)$ suppress small times contributions for $G(t^\prime)$ on the left-hand side of the closed equation, the method assumes the existence of the Laplace transform of the function $G(t)$. Hence, the failure of the method above the critical temperature ($T>T_c$) is interpreted as the failure of the assumption about the existence of the Laplace transform, and $G(t)$ is assumed to increase exponentially at this point, corresponding to an exponential relaxation toward equilibrium. These results accompany those of the reference \cite{Vincent}, which aims to construct a reliable renormalization group for this model. \\

In the second part of this paper, we investigated the validity of our assumptions for the closed equations for $G(t)$, and we study the way the system converges toward equilibrium points using effective dynamics for the quenched variable $a(t)$. We thus elucidate the behavior of $\dot{G}(t)$, and especially the dependency of the previous assumption regarding the order of the zeros of the potential. In this way, we show that for zeros with multiplicity one, the conclusions of the previous section hold, and the system below the critical temperature reaches equilibrium with rate $t^{-1}$, and has power law late time correlations. In the general case, however, the late-time behavior is not the same for odd and even multiplicity. For odd multiplicities, the behavior is essentially the same as for multiplicity one, and after an exponential phase, $G(t)$ behaves as $t^{-3/2}$ for large times. When multiplicity is even in contrast, $G(t)$ diverges exponentially also for zero temperature, and the previous assumption regarding the role of thermal fluctuation is wrong: The system fails to have a power law decays at $T=0$, and due to elementary properties of convolutions, this holds for temperature small enough. One can furthermore provide an estimate of the temperature where the system fails to reach the effective equilibrium point $a(t)=0$. \\

Many of the conclusions of this paper strongly depend on the crude large $N$ limit, and it should be interesting to evaluate finite $N$ effects, regarding for instance the activation of metastable states due to thermal fluctuations (the famous Kramer's problem, see for instance \cite{Chupeau,Burada,Visscher,Berera,Melnikov,Kamenev}. Furthermore, some approximations used in his paper allow only to provide an estimate of transition temperature and should be improved.

\nocite{*}
\vspace{0.5cm}

Data Availability Statement: No Data associated in the manuscript.

\onecolumngrid

\end{document}